\documentclass[12pt,a4]{article}

\usepackage{amsmath}
\usepackage{psfig,epsfig}
\usepackage{amssymb}
\usepackage{graphicx}
\usepackage{subfigure}
\usepackage{cite}
\usepackage{alltt}
\begin{document}

%%%%%%%%%%%%%%%%%%%%%%%%%%%%%%%%%%%%%%%%%%%%%%%%%%%%%%
%%%%%%%%%%%%%%%%%%%%%%%%%%%%%%%%%%%%%%%%%%%%%%%%%%%%%%
\begin{titlepage}

\begin{flushright}
%Ver. 10.2, April 27, 2004\\
% {\bf Preprint no. (*** if any ***) ....} 
\end{flushright}
\vspace{1mm}
\begin{center}{\bf\Large SHARE: Statistical Hadronization\\  
\vspace{2mm}
with Resonances$^{\dag}$}
\end{center}

\vspace{3 mm}

\begin{center}
{\large  G. Torrieri$^{\,a}$,}
{\large  S. Steinke$^{\,a}$,}
{\large  W. Broniowski$^{\,b}$,}
{\large  W. Florkowski$^{\,b,c}$,}\\ 
{\large  J. Letessier$^{\,a,d}$, and}
{\large  J. Rafelski$^{\,a}$}\\

\vspace{0.5cm} 
{\em $^a$ Department of Physics, University of Arizona,
Tucson, AZ~85721}\\ 
{\em $^b$ The H. Niewodnicza\'nski Institute of
Nuclear Physics, \\ Polish Academy of Sciences, PL-31342 Krak\'ow,
Poland}\\ 
{\em $^c$ Institute of Physics, \'Swi\c{e}tokrzyska Academy,
PL-25406 Kielce, Poland}\\ 
{\em $^d$ Laboratoire de Physique
Th\'eorique et Hautes Energies {$^{\ddag}$}\\ Universit\'e Paris 7, 2 place
Jussieu, F--75251 Cedex 05, France.}
\end{center}
\vspace{1mm}
\begin{abstract}
SHARE is a collection of programs designed for the statistical analysis of
particle production in relativistic heavy-ion collisions. With the
physical input of intensive statistical parameters, it generates the 
ratios of particle abundances.  The program includes cascade
decays of all confirmed resonances from the Particle Data
Tables. The complete treatment of these resonances has been known to
be a crucial factor behind the success of the statistical approach. 
An optional feature implemented is a Breit--Wigner type distribution
for strong resonances. An  interface for fitting the parameters
of the model to the experimental data is provided.
\end{abstract}

\begin{center}
{\it Submitted to Computer Physics Communications}
\end{center}

\vspace{1mm}
\footnoterule
\noindent
{\footnotesize
\begin{itemize}
\item[${\dag}$] Work supported in part by  grants from the
U.S. Department of Energy,\\ DE-FG03-95ER40937 and 
DE-FG02-04ER41318, by NATO Science Program,\\
 PST.CLG.979634, and by the Polish State Committee for
Scientific Research\\
 grant 2~P03B 059 25. 
\item[${\ddag}$] LPTHE, Univ.\,Paris 6 et 7
is: Unit\'e mixte de Recherche du CNRS, UMR7589.
\end{itemize}
}

\end{titlepage}

\tableofcontents   

\newpage
%%%%%%%%%%%%%%%%%%%%%%%%%%%%%%%%%%%%%%%%%%%%%%%%%%%%%%%%%%%%%%%%%%%%%%%%%%%%%%%%%

\noindent{\bf PROGRAM SUMMARY}
\vspace{0.3cm}

\noindent{\sl Title of the program:} \- {\tt SHARE}, \hfill July 2004, version 1.1;
\vspace{0.3cm}

\noindent{\sl Computer:}\\
PC, Pentium III, 512MB RAM \hfill \lowercase{NOT HARDWARE DEPENDENT};
\vspace{0.3cm}

\noindent{\sl Operating system:}\\
Linux: RedHat 6.1, 7.2, FEDORA etc.  \hfill \lowercase{NOT  SYSTEM DEPENDENT};
\vspace{0.3cm}

\noindent{\sl Programming language:}\-
{\tt FORTRAN77}: {\tt g77},  {\tt f77}\\
as well as {\tt Mathematica}, 
with temperature-fixed particle yields and 
excluding option of finite width;
\vspace{0.3cm}

\noindent{\sl Size of the package:}\- 
645 KB  directory including example programs
(82KB compressed distribution archive), without libraries (see \\
http://wwwasdoc.web.cern.ch/wwwasdoc/minuit/minmain.html\\
http://wwwasd.web.cern.ch/wwwasd/cernlib.html\\
for details on library requirements  ).
\vspace{0.3cm}

\noindent{\sl Distribution format:}\- 
tar gzip file
\vspace{0.3cm}

\noindent{\sl Number of lines in distributed program, including test data, etc:}\- 
14893
\vspace{0.3cm}

\vspace{0.3cm}

\noindent{\sl Computer:}\- 
Any computer with an f77 compiler
\vspace{0.3cm}

\noindent{\sl Nature of the physical problem:}\\
%---------------------------------------------
Statistical analysis of particle production in relativistic heavy-ion
collisions involves the formation and the subsequent decays of a large number 
of resonances. With the physical input of thermal parameters, such as the 
temperature and fugacities,  and considering  cascading decays, along with weak 
interaction feed-down corrections, the  observed hadron abundances  are obtained.
SHARE incorporates diverse physical approaches, with a flexibility of choice of
the details of the statistical hadronization model, including the selection 
of a chemical (non)equilibrium condition. SHARE also offers  evaluation of the 
extensive properties of the source of particles, such as energy, entropy, 
baryon number, strangeness, as well as the determination of the best intensive 
input parameters fitting a set of  experimental yields. This allows exploration 
of a proposed physical hypothesis about hadron production mechanisms and the 
determination of the  properties of their source.
\vspace{0.3cm}

\noindent{\sl Method of solving the problem:}\\
%---------------------------------
Distributions at freeze-out of both the stable particles and the hadronic 
resonances are set according to a statistical prescription, technically 
calculated via a series of Bessel functions, using CERN library programs. 
We also have the option of including  finite particle widths of the 
resonances. While this is computationally  expensive, it is necessary to 
fully implement the essence of the strong interaction dynamics within the 
statistical hadronization picture. In fact,including finite width has a
considerable effect when modeling directly detectable short-lived resonances
($\Lambda(1520),K^*$, etc.), and is noticeable in fits to experimentally 
measured yields of stable particles. After production, all hadronic 
resonances decay.  Resonance decays are accomplished by addition of the 
parent abundances to the daughter, normalized by the branching ratio. Weak 
interaction decays receive a special treatment, where we introduce 
daughter particle acceptance factors for both strongly interacting decay 
products.  An interface for fitting to experimental particle ratios of the 
statistical model parameters with help of {\tt MINUIT} \cite{minuit} is 
provided. 
\vspace{0.3cm}

The $\chi^2$ function is defined in the standard way. For an investigated 
quantity $f$ and experimental error $\Delta f$,

\begin{equation}
\chi^2=\frac{(f_{\rm experiment}-f_{\rm theory})^2}{(\Delta f_{\rm statistical}
+\Delta f_{\rm systematic})^2}
\label{chi2}
\end{equation}
\begin{equation}
{\rm N_{DoF}}=N_{\rm data\;points}-N_{\rm free\;parameters}.
\end{equation}
(note that systematic and statistical
errors are independent, since the systematic error is not a random 
variable).

Aside of $\chi^2$ the program also calculates the statistical
significance \cite{pdg}, defined as the probability that, given a ``true'' theory and a statistical (Gaussian) experimental error, the fitted $\chi^2$ arises at or above the considered value.
In the case that the best fit has statistical significance significantly below unity, the model under consideration is very likely inappropriate. 
In the limit of many degrees of freedom (${ \rm N_{DoF}}$), the statistical significance function depends  only on $\chi^2/{ \rm N_{DoF}}$, with $90 \%$ statistical significance at $\chi^2/{ \rm N_{DoF}} \sim 1$, and falling steeply at $\chi^2/{ \rm N_{DoF}}>1$.
However, the degrees of freedom in fits involving ratios are generally not sufficient to reach the asymptotic limit.
Hence, statistical significance depends strongly on $\chi^2$ and ${ \rm N_{DoF}}$ separately.   In particular, if ${ \rm N_{DoF}}<20$, often for a fit to have an acceptable
statistical significance, a $\chi^2/{ \rm N_{DoF}}$ significantly less than 1 is required.

The fit routine does not always find the true lowest 
$\chi^2$ minimum. Specifically, multi-parameter fits with too few degrees of 
freedom generally exhibit a non-trivial structure in parameter 
space, with several secondary minima, saddle points, valleys, etc. To help the 
user perform the minimization effectively, we have added tools to compute the
$\chi^2$ contours and profiles. In addition, our program's flexibility allows for 
many strategies in performing  the fit.   It is therefore possible, by following 
the techniques described in Sect. 3.7, to scan the parameter space 
and ensure that the minimum found is the true one.
\vspace{0.3cm}
 Further  systematic 
deviations between the model and experiment can be recognized via the program's output, which includes
a particle-by-particle comparison between experiment and theory.
%-----------------------------------

\noindent{\sl Purpose:}\\
%---------------------------------
In consideration of the wide stream of new data coming out 
from RHIC, there is an on-going activity, with several
groups performing analysis of particle yields.  It is our
hope that SHARE will allow to create an analysis standard 
within the community. It can be useful in analyzing
the experimental data, verifying simple physical assumptions,
evaluating expected yields, as well
as allowing to compare various similar models and programs which are
currently being used. 
\vspace{0.3cm}

\noindent{\sl Computation time survey:}\\
%---------------------------------
We encounter, in the Fortran version computation, times  up to  
 seconds for evaluation  of particle yields. These rise 
by up to a factor of  300 in the process of minimization
and a further factor of a few  when   $\chi^2/{\rm N_{DoF}}$ profiles and
 contours 
with  chemical non-equilibrium are requested. 
\vspace{0.3cm}

\noindent{\sl Accessibility: } 
\vspace{0.3cm}
%-----------------------------------

\noindent {\tt http://www.ifj.edu.pl/Dept4/share.html }\\
or {\tt http://www.physics.arizona/$\tilde{\phantom{~}}$torrieri/SHARE/share.html }\\
%{\tt http://www.physics.arizona/$\tilde{\phantom{~}}$SHARE/share.html }\\
as well as from the authors upon request

\section{Introduction}
%%%%%%%%%%%%%%%%%%%%%%%%%%%%%%%%%%%%%%%%%%%%%%%%%%%%%%%%%%%%%%%%%%%%%%%%%%%%
In strong interaction reaction  processes particle production is abundant. In 
his seminal 1950 article Fermi proposed a non-perturbative description of 
particle yields based on the assumption that the accessible phase space will 
be fully saturated  \cite{Fer50}. Pomeranchuk extended the model by 
reconsidering the particle freeze-out condition \cite{Pom51}. He  argued
that the reaction volume should expand to the point where particles
would decouple from each other, given their inelastic reaction cross 
sections.  We call this stage the chemical freeze-out. In the ensuing decade 
another important feature of the strong interactions was discovered: the existence 
of numerous hadronic resonances. Hagedorn recognized that the large number  of 
different hadronic states is an expression and characterization of their 
strong interactions. He has shown that the increase in the number 
of resonances with their mass has profound implications for the behavior of matter 
at high temperature \cite{Hag65}: hot hadronic matter could undergo a phase 
transition at Hagedorn temperature, $T_{\rm H} \approx 160$ MeV.
Hagedorn referred to this as the boiling point of hadronic matter beyond
which  the gas of quarks would prevail as a new form of matter.
For an introduction into the literature and history of this 
subject see \cite{KMR03}, and for recent developments also \cite{wbwfhag}.

High energy relativistic heavy nuclei (RHI) collisions allow to create,
in the laboratory environment, a fireball of matter at extreme density
and temperature. The immediate objective of this experimental program
is to identify formation of a relatively large volume of deconfined
quark--gluon matter, the quark--gluon plasma (QGP), and to explore its
properties.  This form of matter has existed in the early Universe
just before quarks and gluons evolved into the `normal' hadronic
particles which surround us today. Formation of a region of space in
which quarks can roam freely would confirm that their confinement is a
property of the structured strong interaction vacuum state. For a general
introduction to these physics questions see 
\cite{Letessier:gp}.

Considering the short available lifetime of the fireball, it is
difficult to develop probes capable of uniquely distinguishing a reaction
involving formation of the deconfined QGP state from the one involving
a cascade of reactions between individual confined baryons and mesons.
The characteristic feature of the RHI reaction is the formation of a
large number of hadronic particles. Irrespective if, or not,
deconfinement has been achieved, this constitutes the final state of a
RHI collision. Theoretical considerations suggest that features of the
final hadron abundances are sensitive to the question if deconfined
state has been formed \cite{Koch:1986ud}.

Analysis of the final hadron state can thus lead to important insights 
regarding reaction mechanisms governing the last stage of the QGP evolution.  
We offer here a standardized  version of the Statistical HAdronization
with REsonances (SHARE) approach  allowing to obtain final state hadron particle 
yields. We have at first prepared two platforms, developed in parallel
in the form of Fortran and Mathematica shareware, and verified their input
and logical structure by making sure they yield identical results using the same
input file. However, in physics
applications which require optimization of parameters, Mathematica
turns out to be forbiddingly slow, and Fortran is preferred.
For some other applications, the versatility of Mathematica is of clear
advantage, {\em e.g.,} one can track the feeding from resonances, make
use of the integrated graphics, or modify the notebook to extend 
the model.

Several research groups beginning with P. Koch {\it et al}. \cite{Koch:ij} 
have developed statistical models of hadron production based on the seminal
work by Fermi, Pomeranchuk  and Hagedorn \cite{Fer50,Pom51,Hag65}.  
Hadron yields are computed assuming that hadron
production can be obtained evaluating the accessible phase space
size. Known hadron resonances are incorporated and their decay chain
evaluated. One has to view this type of hadronization as providing a
bottom line, there could be always additional microscopic production
mechanisms which should be most clearly visible for most rarely
produced particles. The experimental data
coming from lower energy collisions of 2--11 GeV at AGS, through the
CERN SPS energies of 8--17 GeV , to the highest presently available
energies at RHIC, reaching 200 GeV, can be understood, sometimes to a
surprising accuracy, with the help of such straightforward statistical
ideas. There are many variants of statistical approaches, with the
common feature that at some point of its evolution the hadronic system
freezes, and the particle fill out the phase space according to
statistical distribution. The particles form a `soup' of both stable
hadrons, as well as hadronic resonances, which later decay, increasing
significantly the yields of stable particles. In addition, the system
expands, {\em i.e.},  undergoes longitudinal and transverse flow, which
is an important phenomenon in heavy-ion collisions, distinguishing
them from elementary hadronic collisions.

We note that statistical hadronization programs
\cite{PBMRHIC,wbwf,wbwfstr,torrieri1,torrieri2,Rafelski:2003ju,Becattini:2003wp}
require a very detailed input of the hadronic spectrum, and
definitions of the subsequent cascading decays of hadron
resonances.  Tacit assumptions can make a difference of physical
significance in the outcome of the analysis.  In addition, some
information is not available for relevant resonances and has to be
assessed by using the current hadron structure knowledge, in
particular regarding particle degeneracy and decay patterns.  We have
excluded from consideration all so called single-$*$ resonances, and
practically all double-$**$ resonances seen in particle data book
\cite{pdg}. The double-$**$ resonances we accept are those where the
discovery record is convincing, but a confirmation experiment not
available. An example are $\Omega$-resonances, see comment on
$\Omega(2470)$ in \cite{pdg}.

The statistical hadronization approach to describe particle yields
arises naturally when considering a pot of boiling quark--gluon soup:
hadrons evaporate with an abundance corresponding to the accessible
phase space. The quark-chemical equilibrium in the pot implies that
the evaporated hadrons are near, but in general not at their chemical
equilibrium. For the dynamically evolving fireball of quark--gluon
plasma, a similar situation arises should the QGP fireball undergo
sudden hadronization in the final stages of its evolution
\cite{Rafelski:2000by}.  In this case QGP breaks rapidly into ever
smaller drops of matter, ultimately consisting of individual
elementary hadronic particle species.  The statistical particle
production model also applies in the opposite limit of a very slow
hadronization process \cite{Koch:1985hk,Braun-Munzinger:2003zd},
assuming that there is time for diverse hadronic particles to be
produced and destroyed in a chemical (re)equilibration process.
%%%%%%%%%%%%%%%%%%%%%%%%%%%%%%%%%%%%%%%%%%%%%%%%%%%%%%%%%%%%%%%%%%%%%%%%%%%%%%%%%%%
\section{Statistical models in a capsule}

The statistical models we are concerned with have the following main
physical ingredients:
\begin{itemize}
\setlength{\itemsep}{-0.1cm}
\item Particle abundances at chemical freeze-out,
\item Resonance decays.
\end{itemize}
For the details of different statistical approaches the reader is
referred to
Refs.~\cite{schne,pbmags,rafacta27,BL,cest,rafacta28,cr,%
PBM99,yg,gazacta,gaznpa,finland1,finland2,wfwbmm,csorgo,zakopane}.

%%%%%%%%%%%%%%%%%%%%%%%%%%%%%%%%%%%
\subsection{Statistical particle distributions}
\label{partdis}

The  densities of particle species $i$ 
are given by the Fermi-Dirac or Bose-Einstein
distribution functions
\begin{eqnarray}
\label{nmdef}
n(m_i,g_i;T,\Upsilon_i)\equiv n_i &=& g_i \int {d^3p \over (2 \pi)^3} 
{1 \over \Upsilon^{-1}_i 
\exp(\sqrt{p^2+m^2_i} / T ) \pm 1} ,\\
&=& \frac{g_i}{2 \pi^2}  \sum_{n=1}^{\infty} (\mp)^{n-1} 
\Upsilon_i^{n} \, \frac{T \, m_i^2}{n} \,
K_2 \left( \frac{n m_i}{T} \right). 
\label{ni}
\end{eqnarray}
The second form, Eq.\,(\ref{ni}), expresses the momentum integrals in
terms of the modified Bessel function $K_2$. This form is practical in
the numerical calculations and is used in the Fortran code. The 
series expansion (sum over $n$) converges when $\Upsilon_ie^{-m_i/T}<1$.
Consideration of this condition is required only for the pion case 
in the range of parameters of interest. In Eq.\,(\ref{ni}), the upper
signs refer to fermions and the lower signs to bosons, respectively.  
$\Upsilon_i$ is the fugacity factor, and $m_i$ is the particle mass. The 
quantity $g_i=(2J_i+1)$ is  the spin degeneracy factor as we distinguish 
all particles according to their electrical charge and mass.  The index $i$ 
labels different particle species, including hadrons which are stable under 
strong interactions (such as pions, kaons, nucleons or hyperons) and hadron 
which are unstable ($\rho$ mesons, $\Delta(1232)$, etc.).

In the most general chemical condition~\footnote{This condition is commonly called  chemical 
non-equilibrium. However,  the conventional equilibrium 
in which existent  particles are redistributed according to chemical 
potentials is maintained here. The  non-equilibrium regarding particle 
production is, in precise terms,  called absolute chemical 
(non)equilibrium.}, the fugacity is defined through the parameters
$\lambda_{I^{\,i}_3}, \lambda_q, \lambda_s, \lambda_c$ (expressing, 
respectively, the isospin, light, strange and charm quark fugacity factors), 
and $\gamma_q,\gamma_s,\gamma_c$ (expressing 
the light, strange and charm quark phase space occupancies, $=1$ for 
absolute yield equilibrium).
The fugacity $\Upsilon_i$ is then  given by:
\begin{equation}
\Upsilon_i=\lambda_{I^i_3} \left(\lambda_q \gamma_q\right)^{N^i_q}
\left(\lambda_s \gamma_s\right)^{N^i_s} 
\left(\lambda_{c} \gamma_{c}\right)^{N^i_{c}}
\left(\lambda_{\bar q} \gamma_{\bar q}\right)^{N^i_{\bar q}}
\left(\lambda_{\bar s} \gamma_{\bar s}\right)^{N^i_{\bar s}}
\left(\lambda_{\bar c} \gamma_{\bar c}\right)^{N^i_{\bar c}}, 
\label{upsilons}
\end{equation}
where
\begin{equation}
\lambda_{q}=\lambda^{-1}_{\bar q},\qquad
\lambda_{s}=\lambda^{-1}_{\bar s},\qquad 
\lambda_{c}=\lambda^{-1}_{\bar c},\qquad 
\label{lambdas}
\end{equation}
and
\begin{equation}
\gamma_{q}=\gamma_{\bar q},\qquad
\gamma_{s}=\gamma_{\bar s},\qquad
\gamma_{c}=\gamma_{\bar c}.
\label{gammas}
\end{equation}
Here, $N^i_q$, $N^i_s$ and $N^i_c$ are the numbers of light $(u,d)$, strange
$(s)$  and charm $(c)$ quarks in the $i$th hadron, and $N^i_{\bar q}$, $N^i_{\bar
s}$  and $N^i_{\bar c}$ are the numbers of the corresponding antiquarks 
in the same hadron.

In the case of the models of Refs.~\cite{PBM99,PBMRHIC,wbwf}, where
the chemical potentials for the conserved quantum numbers are considered
assuming chemical equilibrium ($\gamma_{q}=\gamma_{s}=1$) and absence of charm
($N^i_{c}=N^i_{\bar c}=0$) one has
\begin{eqnarray}
\Upsilon^{\rm \,\,eq}_i=
\exp \left( \frac{B_i \mu_B + S_i \mu_S + I^{\,i}_3 \mu_{I_3}}{T}\right),
\label{upsieq}
\end{eqnarray}
where $B_i$, $S_i$, and $I^{\,i}_3$ are the baryon number,
strangeness, and the third component of the isospin of the $i$th
particle, and $\mu$'s are the corresponding chemical potentials.  In
this case, the two formulations are related by equations:
\begin{equation}
\lambda^{\rm eq}_q = e^{\mu_B/3T}, \,\,\,\,\, 
\label{lameq}
\end{equation}
\begin{equation}
\lambda^{\rm eq}_s = e^{(-3 \mu_s + \mu_B)/3T}, \,\,\,\,\, 
\label{gameq}
\end{equation}
%
%and
\begin{equation}
\lambda^{\rm eq}_{I^i_3} = \lambda_{I^i_3} = 
e^{I^{\,i}_3 \mu_{I_3}/T} .
\label{lami3eq}
\end{equation}
The user is free to input either format for all of the three chemical
potentials ($\mu_B,\mu_S,\mu_{I_3}$) or fugacities
($\lambda_q,\lambda_s,\lambda_{I_3}$) (see section 3.1 for details). All quantities are automatically converted
into fugacities for calculations and fits, however both chemical potentials
and fugacities are presented in the output of the program.

%%%%%%%%%%%%%%%%%%%%%%%%%%%%%%%%%%%%%%%%%%%%%%%%
\subsection{Resonance decays}
\label{resdec}
In the first instance, we consider hadronic resonances as if they 
were particles with a given well defined mass, {\it e.g.,} their decay width
is insignificant. All hadronic resonances decay rapidly after freeze-out, 
feeding the stable particle abundances.  Moreover,  heavy resonances 
may decay in cascades, which are
implemented in the algorithm where all decays proceed sequentially
from the heaviest to lightest particles. As a consequence, the light
particles obtain contributions from the heavier particles, which have
the form 
\begin{eqnarray}
n_1  &=& b_{2\rightarrow 1} 
 \, ... \, b_{N\rightarrow N-1}
n_N,
\label{n1}
\end{eqnarray}
where $b_{k\rightarrow k-1}$ combines the branching ratio for the $k
\rightarrow k-1$ decay (appearing in \cite{pdg}) with the appropriate
Clebsch--Gordan coefficient.  The latter accounts for the isospin symmetry
in strong decays and allows us to treat separately different charged
states of isospin multiplets of particles such as 
nucleons, Deltas,  pions, kaons, {\it etc}. For
example, different isospin multiplet member states of 
$\Delta$ decay according to the following
pattern:
\begin{eqnarray}
&& \Delta^{++} \rightarrow \pi^+ + p, 
\label{D++} \\
&& \Delta^{+} \rightarrow  {1 \over 3} (\pi^+ + n) + {2 \over 3}
(\pi^0 + p), 
\label{D+} \\
&& \Delta^{0} \rightarrow  {1 \over 3} (\pi^- + p) + {2 \over 3}
(\pi^0 + n), 
\label{D0} \\
&& \Delta^{-} \rightarrow  \pi^- + n.  
\label{D-}
\end{eqnarray}
Here, the branching ratio is 1 but the Clebsch--Gordan coefficients
introduce another factor leading to the effective branching ratios
of 1/3 or 2/3, where appropriate.

For two-body strong decays (such as the example shown above), the
Clebsch--Gordan coefficients are automatically calculated by the
program. In this case, the branching ratio taken from \cite{pdg} is
the only input in the program.  For three-body decays, the products of
the Clebsch--Gordan coefficients have been averaged, with equal
weights, over all possible couplings between the three isospins. The
result of this procedure has been used to rescale the branching ratio
taken from \cite{pdg}, and the effective branching ratio has been
obtained in this way, which is delivered as the input.

The case of $\Delta \longrightarrow \pi N$ is easy to deal with, since
only one decay channel is present and the branching ratio is
well known. However, in most of the cases, we take into account
several decay channels appear. We note that the partial widths 
(product of branching ratio with total width) are not
sufficiently well known.  In our approach we disregard, as a rule, all
decays with the branching ratios smaller than 1\%. In addition, if the
decay channels are classified  as {\it dominant}, {\it large, seen}, or
{\it possibly seen,} we always take into account the most important
channel. If two or more channels are described as equally important,
we take all of them with the same weight. For example $f_{0}(980)$
decays into $\pi \pi $ (according to \cite{pdg} this is the {\it
dominant} channel) and $K \overline{K}$ (according to \cite{pdg} this
is the {\it seen} channel). In our approach, according to the rules
stated above, we include only the process $f_{0}(980) \longrightarrow
\pi \pi $. Similarly, $a_{0}(1450)$ has three decay channels: $\eta
\pi $ ({\it seen}), $\pi \eta ^{\prime }(958)$ ({\it seen}), and $K
\overline{K}$ (again {\it seen}). In this case, we include all three
decay channels with the weight (branching ratio)\ 1/3.
Certainly this procedure is not unique and other methods of
selection/treatment of the  decays are conceivable.

Another difficulty is that the branching ratios are not given exactly
(instead of one value, the whole range of acceptable values is given)
and the sum of the branching ratios may differ significantly from
1. In this case we take the mean values of the branching ratios. Since
we require that their sum is properly normalized, sometimes we are
forced to rescale all the mean values in such a way that their sum is
indeed 1.  Of course, a different convention from ours can be
implemented by the user through a different particle data input file.

Weak decays (identified by the fact that they break
strangeness and isospin) are stored separately, and added to the
ratios with a feed-down correction set by the user. However, electromagnetic 
decays such as $\Sigma^0\to \Lambda+\gamma$ are counted as if
it were hadronic in the contribution to the yield of $\Lambda$ (and
the antiparticle). 

%%%%%%%%%%%%%%%%%%%%%%%%%%%%%%%%%%%%%%%%%%%%%%%%%%%%
\subsection{Finite resonance width}
\label{widthdis}

If the particle $i$ has a finite width $\Gamma_i$, the thermal yield of
the particle is more appropriately obtained by weighting Eq.\,(\ref{nmdef})
over a range of masses to take the mass spread into account:
\begin{equation}
\tilde n_i^{\Gamma}=\int\! dM\, 
n(M,g_i;T,\Upsilon_i)\frac{1}{2\pi}\frac{\Gamma_i}{(M-m_i)^2+\Gamma_i^2/4}
\to n_i, \mbox{\ \ for\ \ } \Gamma_i\to 0.
\label{widthn1}
\end{equation}
The use of the  Breit--Wigner  distribution with energy independent width means 
that there is a finite probability that the resonance would be formed at 
unrealistically small masses. Since the weight involves a thermal distribution 
$n(M,g_i;T,\Upsilon_i)$ which would contribute in this unphysical domain, one 
has to use, in Eq.\,(\ref{widthn1}), an energy dependent width. 

The dominant  energy dependence of the width is due to the decay threshold energy 
phase space factor, dependent  on the angular momentum present  in the decay. 
The explicit form can be seen in the corresponding reverse production 
cross sections \cite{phase1,phase2}. The energy dependent partial width in the 
channel $i\to j$  is to a good approximation: 
\begin{equation}
\label{partwid1}
\Gamma_{i\to j}(M)=b_{i\to j}\Gamma_i
     \left[1-\left(\frac{m_{ij}}{M}\right)^2\right]^{l_{ij}+\frac{1}{2}},
          \quad \mbox{for} \quad M>m_{ij}.
\end{equation}
Here, $m_{ij}$ is the threshold of the decay reaction with branching 
ratio $b_{i\to j}$. For example  for the decay 
of $i:=\Delta^{++}$ into $j:=p+\pi^+$, we have $m_{ij}=m_p+m_{\pi^+}$, while the
branching ratio is unity and the angular momentum released in 
decay is $l_{ij}=1$. From these partial widths 
the total energy dependent width arises,
\begin{equation}
\Gamma_i\to \Gamma_i(M) = \sum_{j}\Gamma_{i\to j}(M).
\end{equation}
For a resonance with width, we thus have replacing Eq.\,(\ref{widthn1}):
\begin{equation}
 n_i^{\Gamma} = \frac{1}{N_i} \sum_{j} 
\int_{m_{ij}}^{\infty} dM \, 
n(M,g_i;T,\Upsilon_i)\frac{\Gamma_{i\to j}(M)}{(M-m_i)^2+[\Gamma_i(M)]^2/4},
\label{widthn}
\end{equation}
and the factor $N$ (replacing $2\pi$) ensures the normalization:
\begin{equation}
\label{norm}
N_i= \sum_{j} 
\int_{m_{ij}}^{\infty}  dM 
\frac{\Gamma_{i\to j}(M)}{(M-m_i)^2+[\Gamma_i(M)]^2/4}.
\end{equation}

In principle, Eq.\,(\ref{widthn}) does not take into account the possibility
that the state into which one is decaying is itself a unstable state in 
a thermal bath. Doing this would require a further average over the width 
distribution of receiving state. We will not implement  here this refinement. 

The evaluation of the Breit--Wigner integral requires additional attention.
The presence of the thermal distribution limits  the integral range in 
Eq.\,(\ref{widthn1}) and thus the integral is finite and well behaved. 
However, it involves in general diverse cases of both very narrow, $\Gamma\ll m$, 
and wide, $\Gamma\simeq m$, resonances.
To perform the integral reliably and precisely in all of these cases
we split it in two parts,
$\int_{m_{ij}}^{\infty}= \int_{m_{ij}}^{m_0 + 2 \Gamma} + 
\int_{m_0 + 2 \Gamma}^{\infty}$.
  The first part can be securely done 
using, {\it e.g.,} Gaussian  integration.
The second part, with a potentially slowly falling tail, can be computed through a variable change
$z=\frac{(m_0+2 \Gamma)}{M}$ before the numerical integration.

Calculating widths adds considerably to the computation time.  It should be noted that Eq. \ref{nmdef}
is in the form of $K_2$ functions depending on temperature and mass, weighted by coefficients representing
chemical potentials.   The expensive width integrals, therefore, only have to be recalculated when the temperature under consideration changes.  The program takes advantage of this in its fitting algorithm, something which can dramatically increase fitting time.   See section 3.7 for details.
%%%%%%%%%%%%%%%%%%%%%%%%%%%%%%%%%%%%%%%%%%%%%%%%%%%%%%%%%%%%%%%%%%%%%%%
%%%%%%%%%%%%%   PUBLICATION MODIFICATION   %%%%%%%%%%%%%%%%%%%%%%%
\section{SHARE structure}
\begin{figure}[htp]
\centering
%\vspace{-0.5cm}
  \epsfig{figure=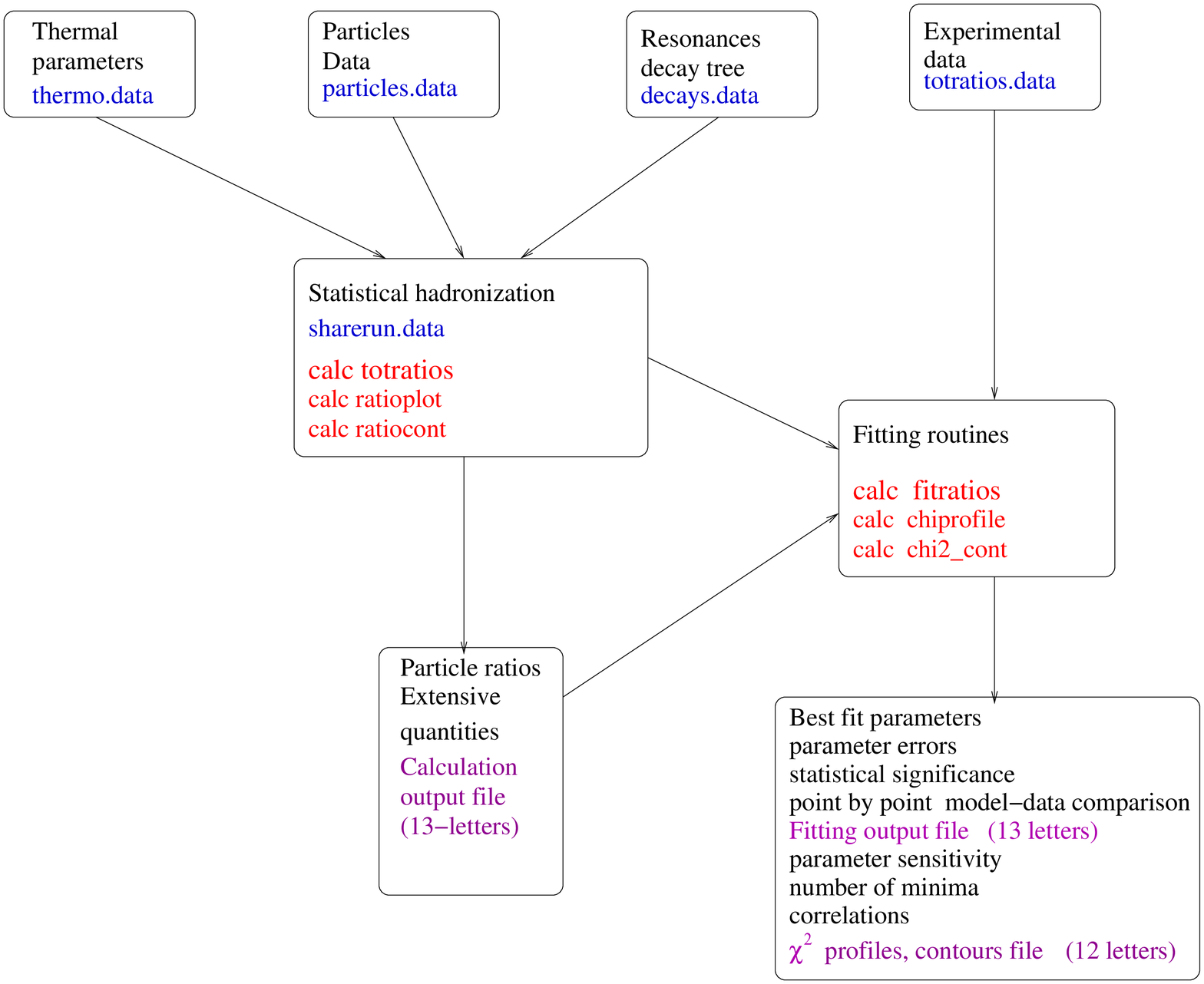,width=14cm}
 \vspace{-0.8cm}
 \caption{Structure of the SHARE package.   Running commands which can be given are in red, default input filenames are in blue, while
possible output files are in violet (color online).   All output filenames are set by the user}
\label{share_structure}
\end{figure}

SHARE is a modular program, whose basic structure is illustrated in Fig. \ref{share_structure}.
Five input data files are required, containing model and experimental data.
There is a calculational and a fitting block, piloted by instructions which the program reads from a running file (called sharerun.data).
Each command will generally be performed independently from the others, and generate its own output in a separate file.
%%%%%%%%%%%%%%%%%%%%%%%%%%%%%%%%%%%%%%%%%%%%%%%%%%%%%%%%%%%%%%%%%%%%%%%%%%%

Any user defined 
file names of specified 
length can be used for all input files except sharerun.data, but the defaults are: \\
\indent {\tt thermo.data} (11 letter filename),\\ 
\indent {\tt particles.data} (14 letter filename),\\
\indent  {\tt decays.data} (11 letter filename),\\
\indent  {\tt totratios.data}  (14 letter filename),\\
\indent {\tt ratioset.data} (13 letter filename).\\

These deal,
respectively, with initial thermal parameters defining the 
beginning of the fit procedure, particle properties,
decay patterns, and the particle ratios 
(Math SHARE uses {\tt particles.data} and {\tt decays.data}
only).  These input files are described in detail in the next
subsections.  It is possible to insert comments into all of these
files: any time the first character starts with `$\#$', the
subsequent input is disregarded (CAUTION: For Math SHARE a comment
line must include at least one more word after `$\#$', {\em e.g.,}
`$\#$ This is a comment' ).

%%%%%%%%%%%%%%%%%%%%%%%%%
\subsection[Thermal parameter input\\
 (default name: thermo.data)]{Thermal parameter input\\
 (11 letter filename, default name: thermo.data)}

The {\tt thermo.data}  file contains input temperature and chemical potentials.
 Where applicable, all units are GeV. A sample is given in Table \ref{thermoini}.
% \vspace{0.5cm}
\begin{table}[h]
\begin{tabular}{lll}
 \textbf{tag} & \textbf{value} & \textit{explanation (not part of file)}\\
 temp &   0.165  & \textit{temperature}\\
 mu$\_$b &   0.028& \textit{light quark fugacity or chemical potential}\\
 mu$\_$s &   0.006& \textit{strange quark fugacity or chemical potential}\\
 gamq  &  1.& \textit{light quark phase space occupancy}\\
 gams  &  1.& \textit{strange quark phase space occupancy}\\
 lmi3  &  -0.001& \textit{$I_3$ fugacity or chemical potential}\\
 norm  &  1.& \textit{absolute normalization}\\
 lamc  &  1.& \textit{charm quark fugacity}\\
 gamc  &  0.001&\textit{charm quark phase space occupancy} \\
 accu  &  0.001&\textit{calculations accuracy}
\end{tabular}
\caption{A typical thermal input file \label{thermoini}}
\end{table} 

\noindent Chemical potentials can be input as $\lambda's$ and  $\gamma's$,
Eqs.\,(\ref{upsilons})--(\ref{gammas}), or  as $\mu's$, 
Eqs.\,(\ref{upsieq})--(\ref{gameq}).  If the tag is, respectively, `mu$\_$b',
`mu$\_$s' or `mui3' the program assumes input is given as in
Eqs.\,(\ref{upsilons})--(\ref{gammas}) otherwise
Eqs.\,(\ref{upsieq})--(\ref{gameq}) are assumed.  

The 4-letter tags given in the file are used throughout the program to 
label the corresponding thermodynamic parameters  (fitting variables, arguments for profiles
and contour plots, {\it etc}).

%%%%%%%%%%%%%%%%%%%%%%%%%
\subsection[Particle properties data file\\  
(default name: particles.data)]%
{Particle properties data file\\
  (14 letter filename, default name: particles.data)}

This input file contains information about the properties of particles
such as: mass, width, spin, isospin, the quark contents, and the
Monte-Carlo identification number.  The data file is written in the
following format

\vspace{0.1cm}

\noindent {\bf name\quad mass\quad  width\quad  spin\quad  I\quad  
I3\quad  q\quad  s\quad  aq\quad  as\quad  c\quad   ac\quad  MC}

\vspace{0.1cm} 

\noindent where:

\begin{description}
\setlength{\itemsep}{-0.1cm}
\setlength{\labelwidth}{1.5cm}
\setlength{\itemindent}{0.5cm}
\item[name] --- a nine-letter character string identifying the particle,
\item[mass] --- mass in GeV,
\item[width] --- width in GeV,
\item[spin] --- spin,
\item[I] --- isospin,
\item[I3] --- 3rd component of isospin,
\item[q,\ s\hfill] --- number of light/strange quarks,
\item[aq,\ as\hfill] --- number of light/strange antiquarks,
\item[c,\ ac\hfill] --- number of charm/anticharm quarks,
\item[MC] --- particle's identification number, usually (where
  applicable) corresponding to the standard Monte-Carlo particle
  identification convention \cite{pdg}.
\end{description}
\noindent For example, the $\Delta(1232)^{++}$ will appear in the
input file as

\vspace{0.1cm}

\noindent {\bf Dl1232plp    1.2320000    0.1200000         1.5      1.5      1.5
  3      0     0      
0        0       0       2224}

\vspace{0.2cm}

\noindent 
The quark number can be a non-integer, to accommodate strong 
interaction flavor mixing such as that of the $\eta$. Note that 
 SHARE calculations are relevant for a strongly interacting system,
 where the relevant states are $K^0$ and $\overline{K^0}$.
$K^0-\overline{K^0}$ mixing is an electroweak process, occurring on a longer
 timescale, and should  be implemented at the end of the calculation.\\

\noindent Several versions of this input file are available, described in detail in section 6.1

\noindent Our particle naming convention is to form a 9-letter name
through a letters-mass-ending combination
 ({\em e.g.,} Lm1115zer for
$\Lambda$), with the following usual three letter endings

\begin{description}
\setlength{\itemsep}{-0.1cm}
\setlength{\labelwidth}{1.5cm}
\setlength{\itemindent}{0.5cm}
\item[zer] for Zero
\item[zrb] for Zerobar
\item[plu] for plus
\item[plb] for plusbar
\item[min] for minus
\item[mnb] for minusbar
\item[plp] for plusplus
\item[ppb] for plusplusbar
\item[sht] for particle with this ending 
sht = $\frac{1}{\sqrt{2}}$ (zer+zrb)\\
{\em e.g.,}  K0492sht for $K_S= \frac{1}{\sqrt{2}} (K_0+\overline{K_0})$
\item[tot] tot  = (zer+zrb)\\
{\em e.g.,}  Lm1115tot for $\Lambda+\overline{\Lambda}$
\item[mnt] mnt  = (min+mnb)\\
{\em e.g.,}  UM1321mnt for $\Omega+\overline{\Omega}$
\item[pmb] pmb  = (plu+plb)\\
{\em e.g.,}  pr0938pmb for $p+\overline{p}$
\item[plm] plm  = (plu+min) \\
{\em e.g.,}  Xi1321plm for $\Xi^+ + \Xi^-$ 
\end{description}
These  conventions will be assumed also  when fitting particle ratios and/or
 yields
%involving  mixed particles
(see section 3.4 for details)

%%%%%%%%%%%%%%%%%%%%%%%%%
\subsection[Particle decay pattern data file\\ (default name: decays.data)]
{Particle decay pattern data file\\
 (11 letter filename,  default name: decays.data)}

This input file contains  the information on particle decays.
The data file lines are here  written in the format:

\vspace{0.3cm}

\noindent {\bf Name$_{\rm parent}$  Name$_{\rm daughter1}$  Name$_{\rm
    daughter2}$  Name$_{\rm daughter3}$\footnote{Where applicable.}  BR
    C--G?(0/1)}

\vspace{0.3cm}

\noindent where {\bf BR} refers to the branching ratio of this decay
(without the Clebsch--Gordan factor, as it appears in the data book)
and C--G refers to whether the branching ratio should be completed by a
Clebsch--Gordan coefficients (0:~no, 1: yes).  For instance, the
decays $\Delta^{+} \rightarrow \pi^+ + n$ and $\Delta^{+} \rightarrow
\pi^0 + p$ will be given as (compare Eqs.\,(\ref{D++})--(\ref{D-})):

\vspace{0.2cm}

\begin{tabular}{lllll}
Dl1232plu  &   pi0139plu   &  ne0939zer   &  1.0    &   1\\
Dl1232plu  &   pi0135zer   &  pr0938plu   &  1.0    &   1\\
\end{tabular}
\vspace{0.2cm}

\noindent while $\eta$ decays will be
 (see discussion in Sect. \ref{resdec}):

\vspace{0.2cm}

\begin{tabular}{lllllc}
eta547zer &  gam000zer &  gam000zer &            &  0.3943 & 0\\
eta547zer &  pi0135zer &  pi0135zer &  pi0135zer &  0.3251 &  0\\
eta547zer &  pi0139plu &  pi0139min &  pi0135zer &  0.226  &  0\\
eta547zer &  pi0139plu &  pi0139min &  gam000zer &  0.0468 &  0\\
\end{tabular}
%\vspace{0.5cm}

%%%%%%%%%%%%%%%%%%%%%%%%%%%%%%%%%%%%%%%%%%%%%%%%%%%%%%%%%%%%%%%%%%%%%%%%%%%
\subsection[(Experimental) Values to be calculated (default: totratios.data)]
{(Experimental) Values to be calculated\\
 (14-letter filename, default file: totratios.data)}
Experimental data values of interest  are submitted 
 in in the following format.  
%%%%%%%%%%%%%%%%%%%%%%%%%%%%%%%%%%%%%%%%%%%%%%%%%%%%%%%%%%%%%%%%%%%%%%%%%%%
\begin{figure}[htp]
\centering
%\vspace{-0.5cm}
  \psfig{figure=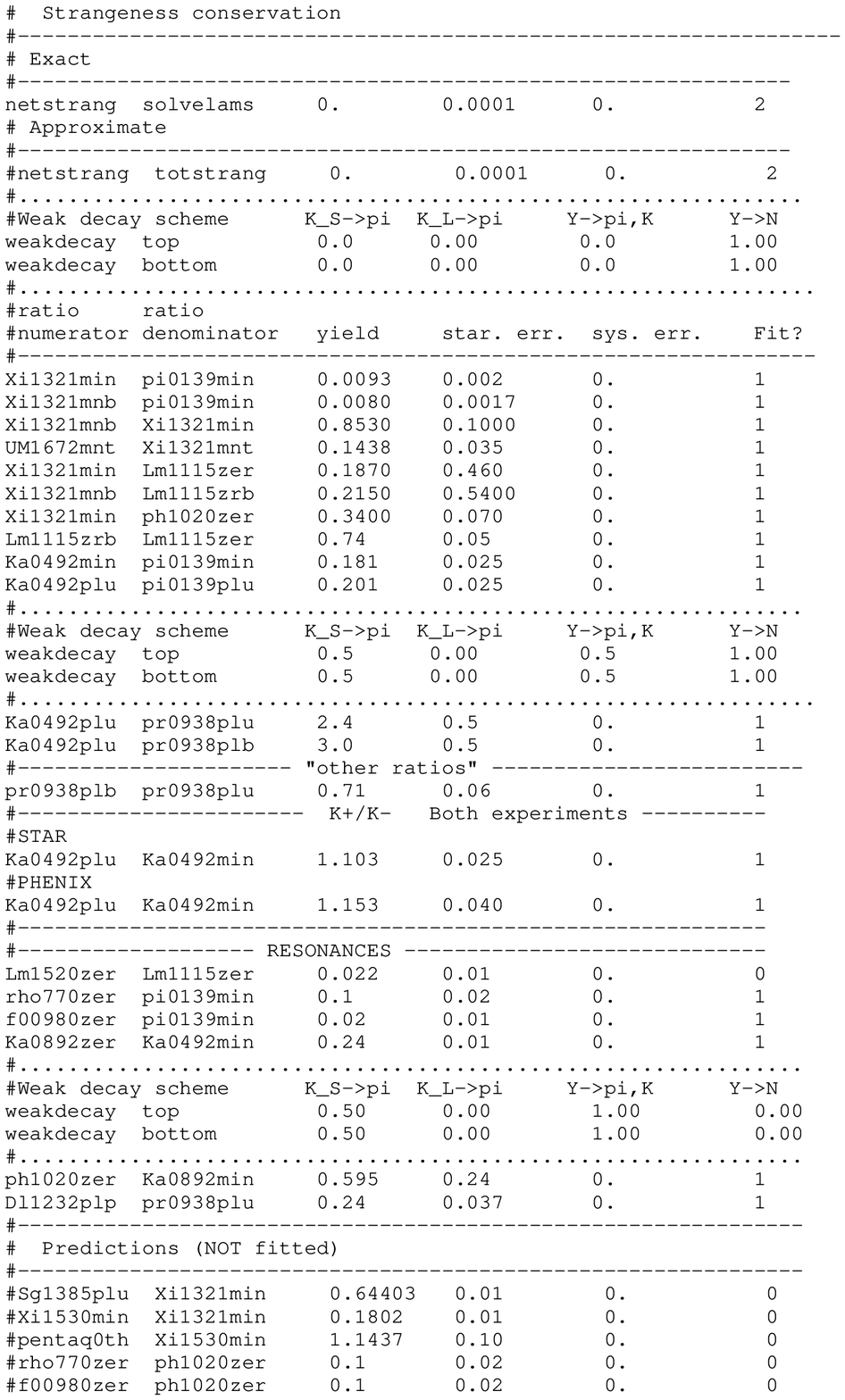,width=16cm}
 \vspace{-0.8cm}
 \caption{A typical totratios.data file}
\label{fig_totratios}
\end{figure}
%%%%%%%%%%%%%%%%%%%%%%%%%%%%%%%%%%%%%%%%%%%%%%%%%%%%%%%%%%%%%%%%%%%%%%%%%%%
%%%%%%%%%%%%%%%%%%%%%%%%%%%%%%%%%
\subsubsection{Particle ratios}
When considering the particle ratio
 evaluation:
\vspace{0.3cm}
%%%%%%%%%%%%%%%%%%%%%%%% PUBLICATION MODIFICATION %%%%%%%%%%%%%%%%%%%%%%%
\noindent {\bf name1  name2    data    random    systematic  fit?($-1$/0/1/2)}

\vspace{0.3cm}

\noindent where

\begin{description}
\setlength{\itemsep}{-0.1cm}
\setlength{\labelwidth}{2.5cm}
\setlength{\itemindent}{1.5cm}
\item[name1] The first particle in the ratio (numerator)
\item[name2] The second particle in the ratio (denominator.  Can also be a tag indicating the quantity is not a ratio but a yield or a density)
\item[data] The experimental value of the data point
\item[random] The random (statistical) error
\item[systematic] The systematic error
% (NB: This program adds the
%random and systematic errors)
\item[fit?] This ratio contributes to the evaluation of
$\chi^2/{\rm N_{DoF}}$ if this parameter is
set to 1 or 2.    If the parameter is set to 0, the ratio is not
fitted, but calculated and output to the graph file
 (see cards in section 4 for details).
if fit = $-1$ or 2 means the ratio is not output to the graph file.  
\end{description}
%%%%%%%%%%%%%%%%%%%%%%%%%%%%%%%%%
\subsubsection{Particle yields and source properties}
\noindent The entries in this files have 
a second use when name1 and name2 are not 
 particle names but are as follows:

In case {\bf name2} is:
\begin{description}
\setlength{\itemsep}{-0.1cm}
\setlength{\labelwidth}{2.2cm}
\setlength{\itemindent}{1.2cm}
\item[prt$\_$yield]  the yield of the first
particle or collective quantity.
\item[prdensity]  the density of the first
particle or collective quantity (in $fm^{-3}$).
\item[solveXXXX] See section 3.6
\end{description}

In case {\bf name1} is:
\begin{description}
\setlength{\itemsep}{-0.1cm}
\setlength{\labelwidth}{2.3cm}
\setlength{\itemindent}{1.3cm}
\item[negatives] All negative particles stable under the strong interaction
\item[totstrang]  
strangeness\footnote{For this and the subsequent quantities, volume normalization can be correct 
only when total particle yields are considered.   The units of the calculated result vary according to
whether the quantity is a density, a yield or a ratio} $\langle s\rangle$
\item[netstrang] net  strangeness $\langle s-\bar s\rangle$
\item[totenergy] energy (in GeV)
\item[totbaryon] sum of all baryons and antibaryons, $B+\overline{B}$
\item[netbaryon] net baryon number, {\em i.e.,}  baryons minus antibaryons, $B-\overline{B}$
\item[totcharge] charge $Q$
\item[netcharge] net charge $<Q-\overline{Q}>$
\item[entropy$\_$t] entropy  $S$
\end{description}
We provide several data files as an example, described in section 6.1
%%%%%%%%%%%%%%%%%%%%%%%% PUBLICATION MODIFICATION %%%%%%%%%%%%%%%%%%%%%%%
\subsubsection{Weak decay feed-downs}
All weak decays are included with the provided SHARE decay tree input files.
They are identified as weak by the program due to the violation of strangeness and isospin within the decay.
Weak decays 
present additional complications, since their reconstruction acceptance
is usually non-trivial and experiment-specific.

SHARE therefore,  includes additional parameters, set by the user, to regulate the acceptance of these decays. 
These parameters are read in from the same file as the experimental data points, and have approximately the same format.

A statement specifying weak decay feed-downs has the form\\
\begin{tabular}{llllll}
  weakdecay  & Tag     &      $F_1$ &  $F_2$ &   $F_3$ &  $F_4$\\
\end{tabular}\\
Where:
\begin{description}
\item[Tag] specifies weather the coefficients refer to the numerator
or denominator of the ratio (note that they are often different, e.g. in a ratio such as $\Xi/h^-$).     If the user is setting the numerator feed-down coefficients, tag is set to ``top''.   In case of the denominator, tag is set to ``bottom''.
\item[$F_{1-4}$] are four numbers, specifying the weak feed-down coefficients.
\begin{itemize}
\item $F_1$ refers to the acceptance of $\pi$ from $K_S \rightarrow \pi \pi$ decays.    It is by default  set to 0.5
\item $F_2$ refers to the acceptance of $\pi$ from $K_L \rightarrow \pi \pi \pi$ decays.   It is typically very small, so it is by default set to 0
\item $F_3$ refers to the acceptance of mesons from weak hyperon decays.
It is by default set to 0.25
\item $F_4$ refers to the acceptance of baryons from weak hyperon decays.
It is by default set to 1.
\end{itemize}
\end{description}
It is possible to specify a different feed-down for each data-point, by preceding a data line with
``weakdecay'' statements, eg:\\
\begin{tabular}{llllll}
  weakdecay  & top     &      1.0 &  0.0 &   0.3 &  0.8\\
  weakdecay  & bottom  &      0.0 &  0.0 &   0.5 &  0.8\\
  Xi1321mnt  & negatives &    0.2 &  0.05 &  0.  &    1
\end{tabular}\\
It is also possible to specify one weak feed-down for a series of data points, or all the datapoints analyzed;   The program assumes all data points after a ``weakdecay'' statement to have the same value until a new ``weakdecay'' statement is encountered.   For example,\\
\begin{tabular}{llllll}
  weakdecay  & top     &      1.0 &  0.0 &   0.3 &  0.8\\
  weakdecay  & bottom  &      0.0 &  0.0 &   0.5 &  0.8\\
  Xi1321mnt  & negatives &    0.2 &  0.05 &  0.  &    1\\
  Lm1115zer  & pr0938plu &    0.15  & 0.05  &  0.1 & 1
\end{tabular}\\
fixes the feed-down for both $\Xi/h^-$ and $\Lambda/p$ ratios.
(In this case, of course, three of the four feed-down coefficients will have no effect on the second data-point)
%%%%%%%%%%%%%%%%%%%%%%%%%%%%%%%%%%%%%%%%%%%
\subsection[Fit parameters (default: ratioset.data)]
{Fit parameters\\ (13 letter filename, default name: ratioset.data)}

The {\tt ratioset.data} file defines the parameters which will be
 varied during the fit or kept constant. It further 
 contains the limits to be used in the fit (for the
use of MINUIT), as well as the initial step size.  The typical format is:

\vspace{0.2cm}
\begin{tabular}{llllc}
 tag  & lower \ /& upper limit &  step size & fit? (0/1) \\[0.1cm]
\hline
 temp &   0.  & 1. & 0.01  &   1\\
 lamq &   0.  & 10.  & 0.1 &   1\\
 lams &   0.  & 10.  & 0.1 &   1\\
 gamq &   0.  & 10.  & 0.01&   1\\
 gams &   0.  & 10.  & 0.1 &   1\\
 mui3 &   0.  & 10.  & 0.2 &   1\\
 norm &   0.  & 10000. & 0.3 &   0\\
 lamc &   0.  & 10.  & 0.1 &   0\\
 gamc &   0.  & 10.  & 0.1 &   0
\end{tabular}
\vspace{0.2cm}

\noindent If the lower and upper limit are equal, MINUIT will fit with no limits. 
If the only fit parameters are particle ratios and densities, the normalization
is automatically kept fixed.   
The parameters can be input in any order.    
We have found that the fit quality (speed, reliability) 
depends considerably on the order of parameter input which is
retained calling the MINUIT package:  firstly, the most significant fit
parameters should be input first (temp, norm). Secondly, the 
highly correlated parameters should be placed next to each other.

Note that values of parameters also arise as result of conservation laws
rather than from fits to particle yields. For example, $\lambda_s$ can be computed
in terms of the other parameters requiring overall strangeness conservation within 
the phase space domain covered by the results considered. $\lambda_{I_3}$ is 
similarly determined via the participant matter proton-neutron ratio, however, 
here it is possible to use this only when the data is available in $4 \pi$ 
acceptance, since other hadrons can buffer the balance of charge condition. 
For example, at RHIC in the central rapidity region the large number of
pions implies in effect nearly symmetric $\lambda_{I_3}=1$. Also, $\mu_B$ can be 
fixed through the participant number in the phase space region observed.
The next section describes how SHARE allows the user to implement
these constraints either by solving the constraints (numerically)
 or by a fit.

%%%%%%%%%%%%%%%%%%%%%%%%%%%%%%%%%%%%%%%%%%%%%%%%%
\subsection{Implementing conservation laws}

SHARE allows the user to solve for a fit parameter, rather than to find its 
value through a fit. For instance, it is possible to implement 
strangeness conservation by solving numerically the constraint equation
\begin{equation}
\langle\, s - \overline{s} \,\rangle = 0
\end{equation}
for $\lambda_s$. In this case, $\lambda_s$ is not a fit parameter 
anymore, but rather an analytical function of the other 
fit parameters and experimental particle yields.

To solve for a fit parameter, \textbf{name2} in the 
totratios.data file should be of the form 
\textbf{solveXXXX} where XXXX corresponds
to the parameter for which one wants to solve.
The parameter limits are still used by the program, 
as a constraint solution outside the limits is rejected.
This is useful for rejecting unphysical solutions of 
the constraint equation, such as $\lambda_s<0$.

It is, in principle, possible to solve any data point for any thermal
parameter.   However, many such combinations do not have minima to
which the equations converge nicely.  If this is the case with
a lot of MINUIT iterations, it is
unlikely that the minimization procedure will work.

It is therefore recommended that the solving algorithm only be used
to solve for chemical potentials from conservation laws.
For instance, to ensure strangeness conservation 
through solving for $\lambda_s$ the input file 
should contain the following line:\\
\noindent {\bf netstrang  solvelams    0.    0.    0.    0.   0.   0}\\
A line solving for $\lambda_q$ from baryon conservation in case of Pb-Pb collisions might be\\
\noindent {\bf netbaryon  solvelamq   362.    0.    0.    0.   0.   0}\\
and the corresponding charge conservation statement will be\\
\noindent {\bf netcharge  solvelmi3   142.    0.    0.    0.   0.   0}\\
Note that it is also possible to implement conservation laws in terms of particle ratios.   The two lines\\
{\bf netbaryon  solvenorm   362.    0.    0.    0.   0.   0}\\
{\bf netcharge  solvelamq   142.    0.    0.    0.   0.   0}\\
will fix $\lambda_q$ in terms of the baryon/charge ($\frac{B}{Q}=\frac{p+n}{p}=0.44$ for Pb-Pb) ratio
even when the absolute normalization \textbf{norm} is a dummy variable which appears in no other
data point and is not used in the fit.

The alternative to exact solving is to implement a conservation law 
by treating it as a data point.
A line such as

\noindent {\bf netbaryon  prt$\_$yield    362.    10.    0.    0.   0.   1}\\
will make sure that the baryon number is close to the one expected for Pb-Pb collisions at SPS. Similarly, 

\noindent {\bf netstrang  totstrang      0.    0.01    0.    0.   0.   1}\\
will make conserve strangeness to one unit in 100 $\overline{s}$ pairs, rather than solving the constraint
exactly.

The choice of whether to implement the conservation law analytically or through 
a fit is therefore left to the user. It was found that both approaches are giving 
very similar results. However the most reliable approach is to use the exact 
constraint only.

{\bf solve} statements should be put at the top of the experimental ratios file.  If this is not done, the program returns with an error.
%%%%%%%%%%%%%%%%%%%%%%%%%%%%%%%%%%%%%%%%%%%%%%%%%%%%%%%
\subsection{Fitting strategies}

Multi-parameter fits can be very time-consuming, 
and occasionally give the wrong answer when a ``false''(i.e.,  local 
rather than absolute)  minimum is found.
If these minima are close enough, MINUIT's error and contour
calculations fail to converge convincingly.    If this is the case, a warning message is
printed in the minimization output file.

The user should make note of  a few simple rules 
which can dramatically decrease the fitting time,
and enhance the probability that the right minimum is found.
\begin{itemize}
\setlength{\itemsep}{-0.1cm}
\item Appropriate calibration of the step size and limits can:
\begin{itemize}
\setlength{\itemsep}{-0.1cm}
\item[---] dramatically decrease the running time for the case that 
the user already knows a good estimate of the minimum,
\item[---] select the appropriate minimum if several minimal points  are 
present in the parameter space. Certain datasets analyzed in the literature present
three minimal points: at equilibrium, at $\gamma_q>1$, and $\gamma_q<1$
\cite{Becattini:2003wp}.
\end{itemize}
\item   Fitting absolute yields is more difficult  (prone to converge to a
false minimal points or spend very long time finding a minimum), as the yield 
normalization strongly correlates with other parameters of the fit.
Especially with a limited experimental  data set  it is not
always easy to distinguish a hot, dense and small fireball
from a colder, more dilute larger one using yields alone.
It is therefore advisable to convert all data points to ratios.
\item Inclusion in the fit  of extensive fireball properties, which require a  
sum over every hadron, considerably increase the running time.  This is 
especially true for energy (density), pressure and entropy (density), which
require the Bessel function evaluation to be redone for every hadron at each 
iteration step. It is therefore suggested that the experimental 
file used for the fit not contain these quantities: They can be calculated after 
the fit in a single pass evaluation, with the fitted parameters as input, using  
modified input files.  The provided experimental input files (see section 6.1)
contain the thermodynamic quantities used in a typical fit (total baryon number, 
strangeness and charge).  We also provide an input file (totratext.data)
devoted exclusively to thermodynamic quantities.
\item The more variable  parameters a fit has, the more minimal points are likely.   In case of
a particularly difficult fit, it is suggested it be completed  in two stages:
An initial fit with only temperature and normalization  as varying parameters, 
followed by a subsequent fit, starting from the previously found minimum, where 
the remaining parameters are released (the sharerun.data provided with the 
program is based on this strategy). Alternatively, the conservation law 
constraints (for e.g., zero net strangeness, given charge and/or baryon number) 
can be tightened or relaxed. A $\chi^2$ profile can also be helpful. Generally, 
a minimum with a $\chi^2$ significantly above expectation signals that the true 
minimum has not been found.
\item Fits with finite resonance widths are much more computationally expensive.   
However, as explained in section 2.3, the computationally expensive integrals and Bessel functions only have to be recalculated when the temperature is changed.
For this reason, a temperature $\chi^2$ profile (which finds the minimum at each value of the temperature parameter, kept fixed throughout each fit.   Hence, the temperature only has to be calculated once per datapoint) can be computed much more quickly than a fit in which temperature is a free parameter.
\end{itemize}
%%%%%%%%%%%%%%%%%%%%%%%%%%%%%%%%%%%%%%%%%%%%%%%%%%%%%%%%%%%
\section{Running program cards (sharerun.data)}
This file contains the instructions which the program executes on running.
Each line corresponds to a different operation, such as reading data files, 
assigning values to parameters, calculating ratios, minimizing, and plotting 
contours and $\chi^2/{\rm N{DoF}}$ profiles. The program will read this file line 
by line, execute each command, and stop when it reaches the end of the file.
Fig.~1 has an example of a typical sharerun.data file. \footnote{The program, 
being written in FORTRAN, is very sensitive to the file's format.   An extra space
may cause the sharerun.data file to be unreadable.   Hence, it is recommended that
the user bases his modifications on the sharerun.data file shown in the 
Fig. 1.}
%%%%%%%%%%%%%%%%%%%%%%%%%%%%%%%%%%%%%%%%%%%%%%%%%%%%%%%%%%%%%%%%%%%%%%%%%%%
\begin{figure}[htp]
\centering
%\vspace{-0.5cm}
  \psfig{figure=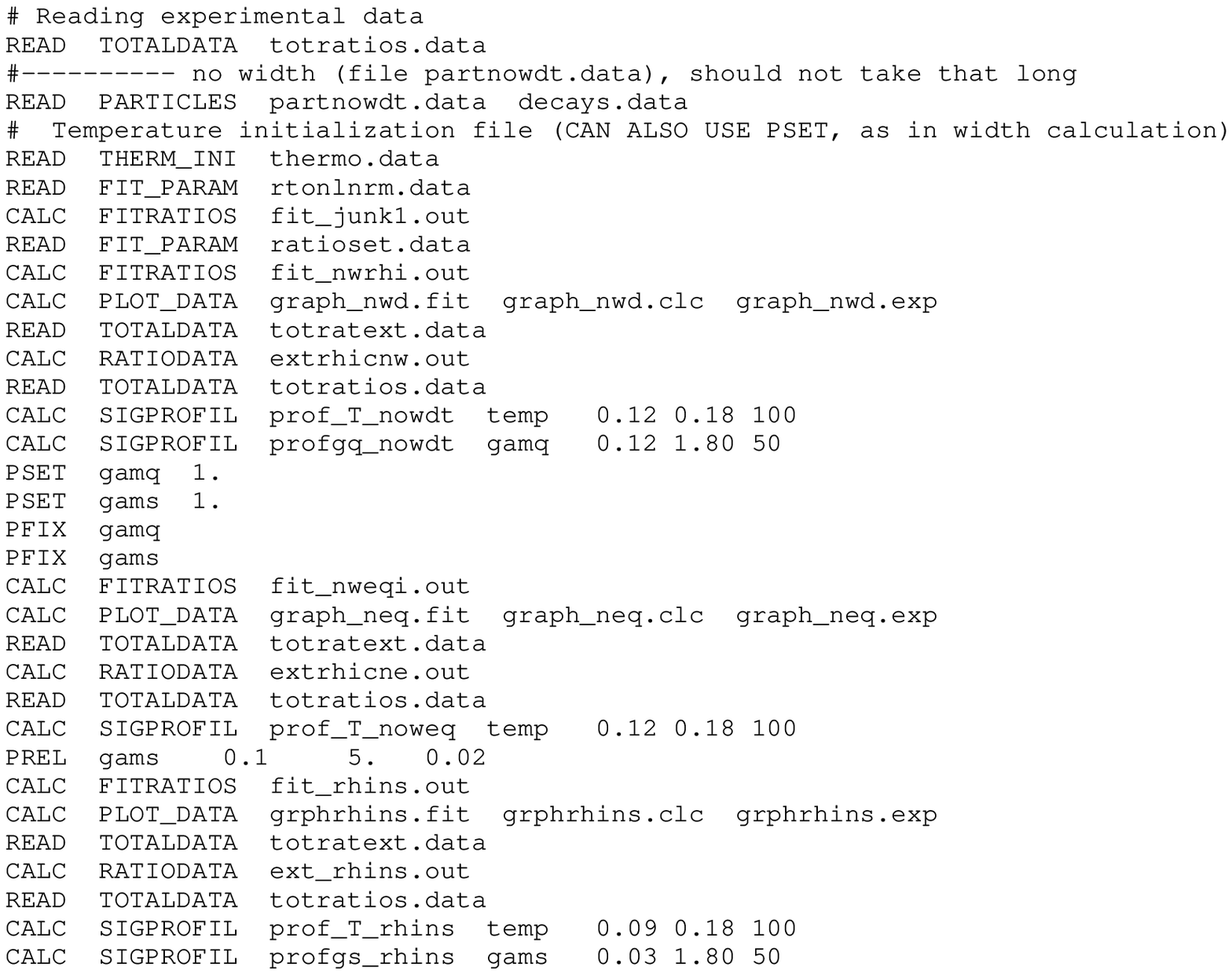,width=16cm}
 \vspace{-0.8cm}
 \caption{A typical sharerun.data file}
\label{sharerunEX}
\end{figure}
%%%%%%%%%%%%%%%%%%%%%%%%%%%%%%%%%%%%%%%%%%%%%%%%%%%%%%%%%%%%%%%%%%%%%%%%%%%

Each command can be used more than
once with different input and output files, one at a time.
We shall proceed to give a detailed description of each 
command's meaning and syntax.  Keep in mind that 2 spaces 
have to be maintained between each word or number. The lines 
have the general form:
\begin{description}
\setlength{\itemsep}{-0.1cm}
\setlength{\labelwidth}{1.5cm}
\setlength{\itemindent}{0.5cm}
\item[PSET $<$4-letter tag$>$ VALUE]\ \\  
This Parameter SET command 
sets the thermodynamic variable defined by the TAG (name)
to its designated VALUE.
The initial command shown in figure \ref{sharerunEX}, 
READ  THERM$\_$INI, comprises 
a series of PSET-type commands which  read from an input file, covering 
all allowed thermal parameters.
\item[READ  THERM$\_$INI  $<$11-letter filename$>$]\ \\  
Reads the file corresponding to
thermo.data, containing the values for the thermal parameters, as described in section 3.1.
\item[READ  PARTICLES  $<$14-letter filename$>$  $<$11-letter filename$>$]\ \\ 
Reads the
file containing particle properties as well as the file 
containing the decay tree, as described in sections 3.2 and 3.3.  
\item[READ  TOTALDATA  $<$14-letter filename$>$]\ \\  Reads the file containing
experimental data and particle ratios to calculate, as described in section 3.4.
\item[PFIX  tag] Fixes the given thermodynamic Parameter to its current value.  In a fit, that parameter will not be a variable.
\item[PREL  tag  $<$Lower limit$>$ $<$Upper limit$>$ $<$Step size$>$  ] Releases the given thermodynamic Parameter, giving the limits and
initial step size used in the fit.   The next command, `READ FIT$\_$PARAM', can
be understood as a series of PFIX and PREL instructions read from an input file.
\item[READ  FIT$\_$PARAM  $<$14-letter filename$>$]\ \\  Reads the file containing
fit parameters, as described in section 3.5.
\item[CALC  RATIODATA  $<$13-letter filename$>$]\ \\  Calculates the value
of the ratios read with the READ RATIODATA command and the current
value of the thermal parameters (obtained either from READ  THERM$\_$INI or a
previous fit).  The output of the calculation is stored with the given
 filename, as a table in
the  following format: \\
  {\bf RATIO NAM1/NAM2 $<$numerical value$>$}
%%%%%%%%%%%%%%%%%%%%  PUBLICATION MODIFICATION  %%%%%%%%%%%%%%%%%%%%%
\item[CALC  RATIOPLOT Datapoint  $<$12-letter Filename$>$  tag L H P]$ $\\
Calculates the ratio of 2 particles, or a particle's yield, or a thermodynamic
quantity, as a function of the variable represented by the 4-letter tag, in a parameter
segment delimited by the limits ({\bf L,H}) and number of points ({\bf P}).
The ratio to be calculated must be in the experimental ratios datafile;  datapoint specifies which point to graph (for instance, if datapoint is 2, the second point from the to will be calculated).
The output is saved as a 2-column table in the given output file, and can be plotted
with packages such as PAW, Mongo or Xmgrace.
%%%%%%%%%%%%%%%%%%%%  PUBLICATION MODIFICATION  %%%%%%%%%%%%%%%%%%%%%
\item[CALC  RATIOCONT  datapoint $<$12-letter Filename$>$]$ $\\{\bf  tag1  L1 H1 P1  tag2 L2  H2  P2}\footnote{one line in the file}\\
Calculates the ratio of 2 particles, or a particle's yield or density, or a thermodynamic
quantity, as a function of two thermodynamic variables represented by the two 4-letter tags, in the parameter
region delimited by the limits ({\bf L1,2  H1,2}) and number of points.  ({\bf P1,2}. Note that an $100\times 100$ grid has 
10\,000 points, and can take a long time to calculate).
The ratio to be calculated is indicated in the same way as in {\bf CALC RATIOPLOT}.
The output is saved as a 3-column table in the given output file, and can be plotted
with a program capable of 3D plots.
\item[CALC  FITRATIOS  $<$13-letter filename$>$]\ \\  Minimizes the $\chi^2/{\rm N_{DoF}}$ of
  the set of experimental data obtained through READ RATIODATA according to
  the parameters in \\READ  FIT$\_$PARAM and initial values in\\ READ  THERM$\_$INI. 
The output is written out to the given filename in the following format: 

  First, the output parameters (+/$-$ error if fitted).

  Then the detailed fit results, as a table with the format:

  {\bf Top Bottom Theory Experiment Tot. error.  Chiterm}

  Where Top and Bottom refer to each ratio's numerator and denominator, and \textbf{Chiterm} for each ratio refers to $\chi$ as defined in Eq.~\ref{chi2}
\begin{equation}
\chi=\frac{f_{\rm experiment}-f_{\rm theory}}{\Delta f_{\rm statistical}+\Delta f_{\rm systematic}} (=0 \mathrm{\; if \; not\; fitted})
\end{equation}
  Finally, the total $\chi^2/{\rm N_{DoF}}$ is presented.  A typical
  output file is shown in Fig.\,2
%%%%%%%%%%%%%%%%%%%%%%%%%%%%%%%%%%%%%%%
\begin{figure}[ht]
%\centering
\hspace{0.8cm}  \psfig{figure=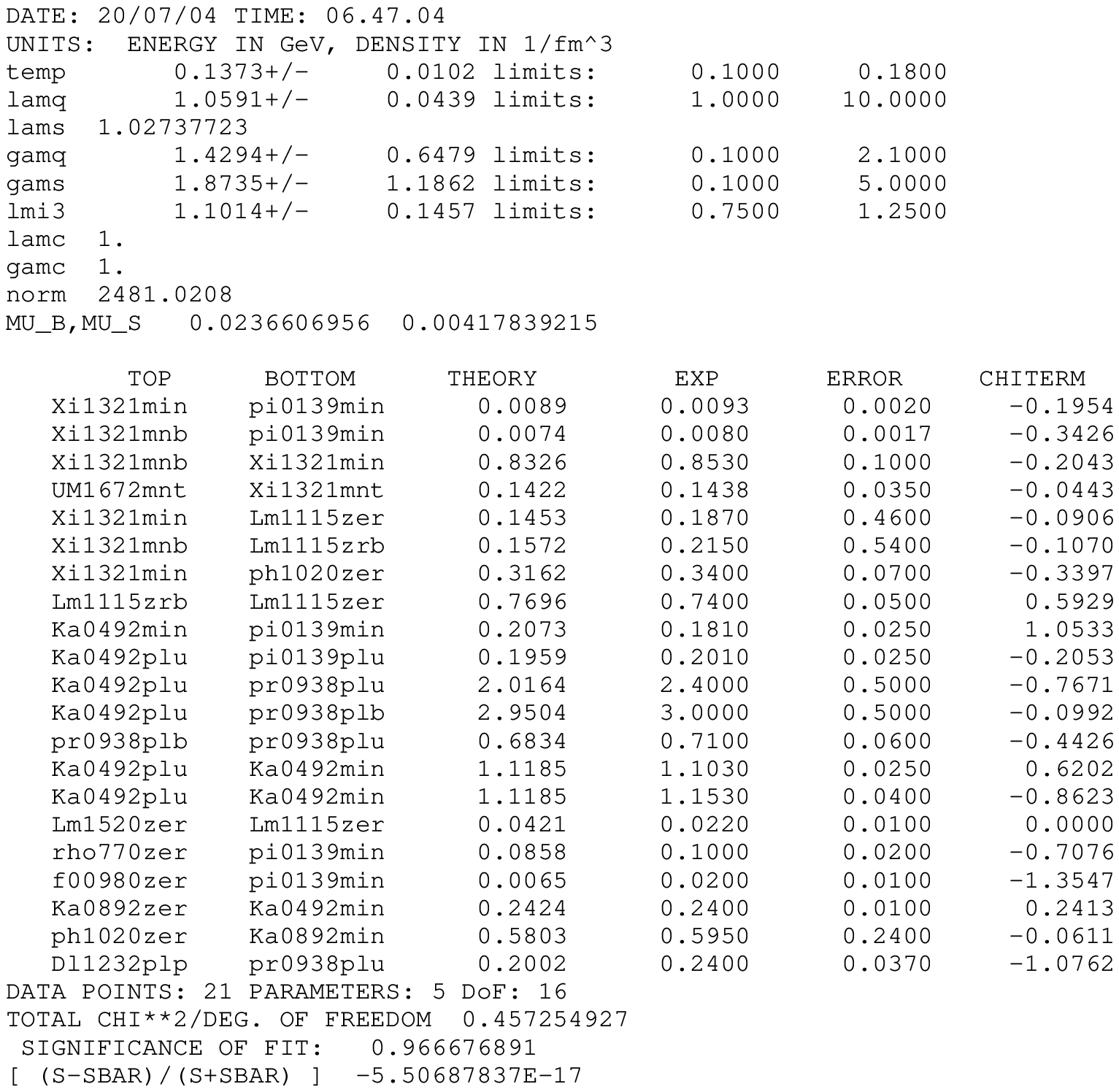,width=16cm}
\vspace*{-3.5cm}
  \caption{A typical fitratios.out file}
\label{fitoutEX}
\end{figure}
%%%%%%%%%%%%%%%%%%%%%%%%%%%%%%%%%%%%%%%
\item[CALC  PLOT$\_$DATA   $<$3 13-letter filenames$>$ ]\ \\ 
Generates three files which are optimized to be graphed by a package such
as PAW, Mongo or Xmgrace.
The first file has a numerical list of ratios which were fitted,
the second, a numerical list of calculated, but not fitted, ratios.
The third has the experimental data, including the error bars.
See the discussion at the bottom of section 3.4
for details on which ratio gets fitted and which just gets calculated.
\item[CALC  CHIPROFIL  $<$12-letter file$>$  tag  L  H  P]\ \\  
This computes a $\chi^2/{\rm N_{DoF}}$ profile of the Parameter designated by tag $<$Parameter tag$>$
(see the thermal input file for a list), from the L to the
  H limit (real numbers), with P specifying the number of computed points (integer). The given file will store the main result, as a 2 column table of the
  parameter value and $\chi^2/{\rm N_{DoF}}$.  The minimum of each of the other
  parameters for each data point will also be stored in files in which
  the parameter is appended to the name.  For instance, if the
  T, $\chi^2/{\rm N_{DoF}}$ profile is stored in file 'profTforratio' the minimal points of
  $\gamma_q$  across the $\chi^2/{\rm N_{DoF}}$ profile are stored in
  file `profTforratio$\_$gamq'.
%%%%%%%%%%%%%%%%%%%%  PUBLICATION MODIFICATION  %%%%%%%%%%%%%%%%%%%%%
\item[CALC  SIGPROFIL] This command is very similar to\\ CALC CHIPROFIL.
However, instead of the $\chi^2$,  the profile for the statistical significance is calculated.   The command syntax is identical to CALC  CHIPROFIL
\item[CALC  CHI2$\_$CONT  $<$9-letter filename$>$  deviation tag1  tag2 ] 
  Computes a $\chi^2/{\rm N_{DoF}}$ contour of parameters denoted by tag1 and
  tag2, with a given deviation from the $\chi^2/{\rm DoF}$ minimum ({\it e.g.,} 
1.1 for a $1.1\times\chi^2_{min}$ contour).  The contour is stored in an output file 9 characters long.
\end{description}
\subsection{Run log (sharerun.out)}
The complete `log' for each run is saved in the file sharerun.out.
This includes:
\begin{itemize}
\setlength{\itemsep}{-0.1cm}
\item The content of each input file (in the same format as read)
\item A list of performed operations
\item The output from MINUIT
\item The content of each output file (in the same format as in the file)
\end{itemize}
If the program ends without a problem, the message `PROGRAM TERMINATED
SUCCESSFULLY' is printed, both on the screen and this file.
If an error occurs, the program writes to the screen that an error
has occurred, and outputs the details of the error to sharerun.log.
%%%%%%%%%%%%%%%%%%%%%%%%%%%%%%%%%%%%%%%%%%%%%%%%%%%%%%%%%%%%%%%%%%%%%%%%%%%%%%
\section{Installation}
\label{sec:installation}

{\tt SHARE} is distributed in a form of an archive containing source and data files.
Running under the Linux operating system is supported presently. We have checked {\tt SHARE}
on FORTRAN  in a few Linux installations.
In order to run {\tt SHARE} one needs:
\begin{itemize}
\setlength{\itemsep}{-0.1cm}
    \item {\tt FORTRAN} compiler e.g.  g77, f77;
    \item {\tt CERN} library \\SHARE needs the CERN Library, 
in particular MINUIT and  the CERN mathematical
routines library. These can be found on  the CERN website if not
installed in current environment.
\end{itemize}
After unpacking, one needs to compile {\tt SHARE} using the included
shellscript `fortrat' (which needs to be amended with the correct path
to where the CERN libraries are)

%%%%%%%%%%%%%%%%%%%%%%%%%%%%%%%%%%%%%%%%%%%%%%%%%%%%%%%%%%%%%%%%%%%%%%%%%%%%%%
%%%%%%%%%%%%%%%%%%%%%%%%%%%%%%%%%%%%%%%%%%%%%%%%%%%%%%%%%%%%%%%%%%%%%%%%%%%%%%
\section{Organization of the Fortran package}
\label{sec:organization}
%%%%%%%%%%%%%%%%%%%%%%%%%%%%%%%%%%%%%%%%%%%%%%%%%%%%%%%%%%%%%%%%%%%%%%%%%%%%%%
%%%%%%%%%%%%%%%%%%%%%%%%%%%%%%%%%%%%%%%%%%%%%%%%%%%%%%%%%%%%%%%%%%%%%%%%%%%%%%
\subsection{Directory tree}
\label{dir-struct}
%%%%%%%%%%%%%%%%%%%%%%%%%%%%%%%%%%%%%%%%%%%%%%%%%%%%%%%%%%%%%%%%%%%%%%%%%%%%%%
\begin{description}
\setlength{\labelwidth}{1.7cm}
\setlength{\itemindent}{0.7cm}
\item [\underline{source code:}] 
\item [sharev1.1.f] {\tt F77}
\item [share1.0.nb] {\tt mathematica}
\setlength{\labelwidth}{1.7cm}
\setlength{\itemindent}{0.7cm}
\item  [\underline{Data files:}]
\item  [thermo.data] Thermal inputs
\item  [particles.data]  Particle datafile --- full width
\item  [partnowdt.data]  Particle datafile --- no width
\item  [decays.data]  Full decays file
\item  [dec$\_$no.data]  Empty decays file
\item  [ratioset.data]  Minimization settings
\item  [rtonlnrm.data]  Minimization settings with most parameters fixed (used in the course of an intermediate minimization,
see section 3.7)
\item  [totratios.data] Sample input drawn from RHIC data (2004).
\item  [totratext.data] Sample input with extensive thermodynamic functions
\item  [chi2.data] Sample input in  Mathematica-readable format
\item  [sharerun.data] running script optimized for RHIC
\item  [\underline{Shell-scripts:}]
\item  [fortrat]  Sample compiling shell-script
\item  [tarrat] Sample archiving shell-script
\end{description}
%%%%%%%%%%%%%%%%%%%%%%%% PUBLICATION MODIFICATION %%%%%%%%%%%%%%%%%%%%%%%
\section{SHARE webpage edition}
A reduced version of SHARE is also accessible directly from the web, at the same website from which the program is downloaded.
It is possible to submit thermal parameters, and calculate a select table of experimentally relevant ratios and thermodynamic quantities.    
The webpage can be used as a cross-check between statistical hadronization packages, and for a comparison with experimental data.  
For full input and calculational versatility, however, we recommend downloading the full version
%%%%%%%%%%%%%%%%%%%%%%%%%%%%%%%%%%%%%%%%%%%%%%%%%%%%%%%%%%%%%%%%%%%%%%%%%%%
\section{Mathematica version of  SHARE}

The Mathematica notebook Math SHARE, included in 
the SHARE distribution, has a similar functionality as 
the Fortran program but greatly restricted physical scope. 
It uses the same universal input files as the Fortran, namely 
{\tt particles.data}, {\tt decays.data}, and in addition the file {\tt chi2.data}.
The notebook is designed to perform three types of calculations:  
\begin{enumerate}
\setlength{\itemsep}{-0.1cm}
\item Run the thermal model for specific thermal parameters $T$,
$\mu_B$, $\mu_S$, and $\mu_I$ (in the Mathematica code the $\gamma$
factors of Sect. \ref{partdis} are set to unity) and determine the
particle ratios,
\item Fit the thermal parameters to the experimental ratios read from
file {\tt chi2.data} via the $\chi^2$ method,
\item Track the feeding of a specified particle from decays of higher
states. This shows the physical importance of high-lying states and
allows to determine the amount of feeding for a given state.
\end{enumerate}

The installation of Math SHARE requires only setting the path to 
the data files (the first cell of the notebook).
The following modules are particularly useful when running the notebook:

\begin{itemize}
\setlength{\itemsep}{-0.1cm}
\item {\tt readpart[fn]} reads the particle properties from the file {\tt fn}
and initializes the primordial yields to zero
\item {\tt decay[fn1]}  performs the decays according to the file {\tt fn1}
(CAUTION: these to modules should be always run in sequence, {\em i.e.},\\ 
{\tt 
... \\
readpart[fn] \\
decay[fn1] \\
...
}
\item {\tt prop[part]} displays the properties of particle {\tt part}

\item {\tt prrat[part1,part2]} gives the ratio of yields of particles {\tt part1}
and {\tt part2} after the resonances have decayed

\item {\tt prrat0[part1,part2]} gives the ratio of `primordial' 
(before the decays) yields of particles {\tt part1} and {\tt part2} 

\end{itemize}

The following flags are used throughout the notebook to control the output and 
decays:

\begin{itemize}
\setlength{\itemsep}{-0.1cm}
\item {\tt listp} (False or True) -- when True, echos the properties of read
particles in the {\tt readpart} module  
\item {\tt describe} (False or True) -- controls the output of the {\tt readpart} and 
{\tt decay} modules. It is recommended to set it to True when running for the first time  

\item {\tt debug} (False or True) -- when True, tests of the input files are performed
while running the {\tt readpart} and 
{\tt decay} modules. In particular, a check is made if the decays proceed from the heaviest to the
lightest particle (otherwise the DECAY OUT OF SEQUENCE warning is printed), the conservation 
of the baryon number and the electric charge is checked, and finally, the sum of the branching
ratios is printed (of course, it should be equal to 1 within the assumed accuracy).
It is recommended to set {\tt debug} to True when running after modification of the 
input files.  

\item {\tt yes3b} (True or False) -- when True, three-body decays are included. This is the 
normal mode. Switching off the three-body decays allows to test their importance

\item {\tt nomass} (True or False) -- when True, decays into product whose total mass is heavier
than the mass of the decaying particle are allowed (!). There are a few decays like this 
in the Particle Data Tables, where the decaying particle is sufficiently wide. To switch off
such decays, set {\tt nomass} to False

\item {\tt threshmass} (Infinity or number) -- when Infinity, 
all decays are included, when set to a number, only particle lighter than threshmass are 
decayed

\item {\tt tracking} (False or True) -- when True, tracking of feeding to 
particle {\tt tracklab} is made

\end{itemize}

For convenience, particle labels {\tt negatives} and {\tt positives} are introduced. They
include all negative or positive charge ground-state 
hadrons which are stable with respect to the strong interactions. 

Furthermore, we include the
weak-decays feeding corrections as done by the STAR collaboration at RHIC. The STAR 
procedure excludes the feeding of pions from the decays of $\Lambda(1115)$. We 
define labels of such pions as {\tt pi0139pluSTAR} or {\tt pi0139minSTAR}
at the end of the {\tt decay} module. Similar corrections can be defined for other 
ways of including the weak-decay corrections.
 
The format of the {\tt chi2.data} file, containing the experimental information 
for the fitting, is \\
{\tt part1 part2 ratio stat.err syst.err}

\noindent More information on how to run Math SHARE is included in comments in the notebook itself.

%\section{Outlook}
%\label{sec:outlook}
%In the future we plan to
%\begin{itemize}
%\item Implement resonances with finite with (this is actually allready
%done in the `beta-version', but temporarily disabled as we are still
%testing the program)
%\item Implement particle spectra, $v_2$ and HBT
%\end{itemize}

%%%%%%%%%%%%%%%%%%%%%%%%%%%%%%%%%%%%%%%%%%%%%%%%%%%%%%%%%%%%%%%%%%%%%%%%%%%
\section{Conclusion}

We hope that SHARE will be useful both as an 
auxiliary tool in the 
analysis of experimental data and as a benchmark to verify  
more detailed ideas about the statistical approach to particle production in 
relativistic heavy-ion collisions. Also, the modules with the particle 
data and the resonance decays may be useful 
in their own right and interfaced to other models of particle production. 
We hope that given the  versatility and modular structure of SHARE it
will  become a useful widely applied tool in study of hadron production in
 heavy-ion physics.

\section*{Acknowledgments}
SHARE builds on our long-term experience in the study of hadronic
particle production.  We thank all our collaborators who over the
years made important contributions to the field.

% We thank ...  (*** if there is anyone ***) 
% for numerous discussions and valuable remarks.

\end{document}